\documentclass{aastex}          
\usepackage{spr-astr-addons}    

\begin{document}
\title{The rms-flux relations in different branches in Cyg X-2}

\shorttitle{rms-flux relations in Cyg X-2} \shortauthors{Z.B. Li et
al.}

\author{Z.B. Li\altaffilmark{1}}
\affil{lizhibing@xao.ac.cn}
\and
\author{L.M. Song\altaffilmark{2}}
\and
\author{J.L. Qu\altaffilmark{2}}
\and
\author{Y.J. Lei\altaffilmark{3}}
\and
\author{J.Y. Nie\altaffilmark{2}}
\and
\author{C.M. Zhang\altaffilmark{3}}

\altaffiltext{1}{Xinjiang Astronomical Observatory, Chinese Academy
of Sciences, 150, Science 1-Street, Urumqi, Xinjiang 830011, China}
\altaffiltext{2}{Key Laboratory for Particle Astrophysics, Institute of High Energy Physics, Chinese Academy of Sciences, 19B YuQuan Road, \\
Beijing, 100049, China} \altaffiltext{3}{National Astronomical
Observatories, Chinese Academy of Sciences, Bejing, 100012, China}

\begin{abstract}
In this paper, the rms-flux (root mean square-flux) relation along
the Z-track of the bright Z-Source Cyg X-2 is analyzed using the
observational data of Rossi X-ray Timing Explorer (RXTE). Three
types of rms-flux relations, i.e. positive, negative, and
'arch'-like correlations£¬ are found in different branches. The rms
is positively correlated with flux in normal branch (NB), but
anti-correlated in the vertical horizontal branch (VHB). The
rms-flux relation shows an 'arch'-like shape in the horizontal
branch (HB). We also try to explain this phenomenon using existing
models.
\end{abstract}

\keywords{stars: individual (Cyg X-2)--stars: neutron--X-rays:
stars}

%
\section{Introduction}
\label{sec:intro} Aperiodic X-ray variability is a common property
in X-ray binaries (XRBs) and active galactic nuclei (AGN)
\citep{b29,b32,b17}. Many works have been done to investigate the
aperiodic X-ray variability. For example, a linear correlation
between rms (see Equation 1) and flux of the X-ray in
AGN and XRBs is reported by \citet{b26}. This relation is found in
both neutron star XRBs (NSXRBs) and black hole XRBs (BHXRBs)
\citep{b26,b15,b6,b10}. Accordingly, the linear relation is
supposed to be a fundamental property of accretion flows in compact
objects. Recently, \citet{b9} report that this linear relation also
exists in a Ultraluminous X-ray source, NGC 5408 X-1, which further
strengthens the above supposition.

The rms-flux relation is very useful in evaluating X-ray variability
models. Considering that the shot-noise model predicts a stationary
power spectrum and can't produce a linear rms-flux relation,
\citet{b26} pointed out that this model is not correct. Furthermore,
\citet{b27} showed that in the accreting millisecond pulsar SAX
J1808.4-3658 is coupled with the 401-Hz pulsation. This indicates
that the linear relation is not produced in the corona but in the
accretion flow onto the magnetic caps of the neutron star (NS),
which means that the linear relation does not favor the coronal
flare model for the X-ray variability neither. Based on the
investigation of the rms-flux linear relation on different
time-scales, \citet{b28} ruled out several X-ray variability models
except the perturbed accreting-flow model (e.g. density waves,
\cite{b18}) of \citet{b16}. On the other hand, \citet{b36} found a
similar rms-flux relation in the Sun and $\gamma$-ray bursts,
suggesting that the accreting-flow model may be also wrong for the
X-ray emission since the Sun does not have accretion flows. In
conclusion, the rms-flux relation is very important in
distinguishing X-ray variability models.

\citet{b6} investigated the rms-flux relation in all known spectral
states of a BHXRB, Cyg X-1, and found that the
linear relation exists in all spectral states although their power
density spectra (PDS) show rather different shapes \citep{b6}.
This result implies that the rms-flux relation is a more universal
characteristic than PDS shape.

Cyg X-2 is a bright, persistent low mass XRB (LMXB) \citep{b1}. It
contains a NS \citep{b11}, with an optically measured mass of
$\sim$1.78$\pm$0.23$M_{\odot}$ and distance of 7.2$\pm$1.1 kpc
\citep{b22}. The orbital period is $\sim$9.8 days \citep{b5,b2}.
For comparing the differences between a BHXRB and a NSXRB, we will
investigate the rms-flux relations of different states in Cyg X-2 to
check out whether the linear relation is valid or not throughout all
the states of a NSXRB.

Cyg X-2 is a Z source according to its track through different
spectral states on the color-color diagram (CCD)
\citep{b7,b8,b30}. In general, a Z source has three branches, i.e.
the horizontal branch (HB), normal branch (NB) and flaring branch
(FB). The HB usually locates at the top of the CCD, followed by the
NB crossing the diagonal, and the so-called FB lies at the bottom of
the CCD. Usually, a Z source can trace out a Z-track in few hours,
but the Z-track will shift in weeks or months. Sometimes, the HB may
have two portions: one is horizontal and the other is vertical. We
call this vertical portion vertical horizontal branch (VHB)
\citep{b24,b12,b25}. During the evolution of a Z-source, it moves
continuously on the Z-track, due to different mass accretion rate
\citep{b30}. From simultaneous multi-band observations, it is
thought that the mass accretion rate reaches a minimum at the left
end of the HB, increases as the source travels from HB, throughout
NB, to FB, and reaches its maximum at the right end of the FB
\citep{b7,b8}. However, \citet{b4} has a different opinion about
the mass accretion rate (see \cite{b4} and reference therein).

Cyg X-2 varies on time-scales from milliseconds to months
\citep{b13,b35}. Several kinds of quasi-periodic oscillations
(QPOs) are detected in Cyg X-2 and its rapid variability is also
found to be related with its position in the CCD. The QPOs that
appear in the HB and the upper NB are called HB oscillations with
the centroid frequency varying from $\sim$ 10 Hz to $\sim$ 70 Hz.
The NB oscillations locate in the lower NB with the centroid
frequency at $\sim$ 5-7 Hz \citep{b7,b30,b23,b31}. In addition,
some kHz QPOs are observed in Cyg X-2 when the source moves from the
left end of the HB to the upper NB (\cite{b34} and references
therein).

Section 2 is data reduction. The rms-flux relations are given in
Section 3. Discussions are presented in Section 4.

\section{Observations and data analysis}
The observational data we used here are from RXTE archive, and are
reduced by standard Heasoft package released by HEASARC.

\subsection{Observations}
In order to investigate the rms-flux relation in different branches,
we require the exposure time as long as possible. Taking into
account that the time interval Cyg X-2 stays in only one branch is
$\sim$ 10 000 s, we use only those observations whose duration is
longer than 10 000 s. Moreover, we discard those observations with
different number of PCUs switched on during the observation (see \it
{The RXTE Cook Book}\rm) and also those observations locate at the
FB or at the lower part of NB, since their PDS always have negative
Power Density in the frequency range we are interested in. We list
the observation IDs of the satisfactory observations in Table
\ref{ObsIDs}.

The first 7 observations belong to Epoch II and the other are during
Epoch III. Their durations are longer than 10ks. In ObsIDs
10065-02-01-000 and 10065-02-01-001, the event mode data,
E\_125us\_64M\_0\_1s, are used for investigation while the binned
mode data, B\_2ms\_16A\_0\_35\_Q, are used in the other 14
observations. We also list the location by their CCD position and
PDS.

The CCDs of each observation are made by using the Standard 2 mode
data with time resolution 16 s. We define the soft color as the
count-rate ratio between 4.7-7.5 keV and 2.0-4.7 keV, and the hard
color between 11.2-18.8 keV and 7.5-11.2 keV during Epoch II (see
also \cite{b14}). Because the PCU gains are different for different
Epoches (see \it {The RXTE Cook Book}\rm), the soft and hard colors
of Epoch III are defined a little different from that of Epoch II,
i.e., the count-rate ratio between 4.1-6.2 keV and 2.0-4.1 keV, and
between 9.4-15.9 keV and 6.2-9.4 keV for soft and hard colors
respectively. However, the difference is corrected using the Crab
observations. All the above count rates are background-subtracted.
The PDS of all the observations are extracted from the Binned Mode
or Event Encoded mode (see Table \ref{ObsIDs}) to locate their
positions in CCD. The branches of those observations are listed in
Table \ref{ObsIDs}.

\subsection{rms calculation}
The Binned Mode or Event Encoded mode data are also used to obtain
the rms-flux relations. For each observation, First, we extract the
4 ms time bin resolution light curve of Channel 0-35 (corresponding
to 2-11.2 keV, 2-13 keV for Epoch II and Epoch III respectively).
Second, we divide the light curve into 32s segments and sort them by
their count rate. Third, we average these segments into several flux
bins in order of the count rate with each flux bin contains $\sim$
20 segments (Actually the rms-flux relation does not vary with the
time-scale of the flux bin, this is also consistent with the result
in Cyg X-1 that the linear rms-flux relation is valid not only on
short time-scales but also on longer time-scales, see \cite{b6}). Fourth, 
we produce the PDSs of
flux bins using the program 'POWSPEC 1.0' with the Poisson noise
subtracted and the power normalized to squared mean intensity
(\citep{b20}). Finally, we integrate the power in a certain
frequency range and multiply the square root of the integral value
by the flux (count rate) of the flux bin, i.e.,
\begin{equation}
{\sigma}={\left\{\int_{f_{1}}^{f_{2}}\langle P(f)\rangle
df\right\}^{1/2} F},
\end{equation}

where $\sigma$ means the absolute rms, $\langle P(f)\rangle$ is the
average power of a flux bin, which can read out from the PDS, and F
is the count rate of the flux bin (normalized to one PCU), $f_{1}$
and $f_{2}$ are the lower and upper integral limits fixed at 0.125
Hz and 60 Hz respectively so as to include the contribution of QPOs.
In this way, we can get the rms and flux of all the flux bins of an
observation and thus derive the rms-flux relation.

The error of rms is given by the following equation:
\begin{equation}
(\Delta \sigma)^{2}=(\frac{\partial \sigma}{\partial P(f)})^{2}
(\Delta P(f))^{2}+(\frac{\partial \sigma}{\partial F})^{2} (\Delta
F)^{2}
\end{equation}
where $\Delta P(f)$ is error of P(f), and $\Delta F$ is the error of
the flux. As usual, we use the count rate of each flux bin as the
flux (\citep{b26,b6,b15,b9}).

In this paper we have not corrected the effect of dead time of the
instruments, because it accounts for only $\sim$1.2\% of the total
rms and doesn't affect our results. (we use the dead time model of
\cite{b37} to compute the instrumental dead time modification. See
also equation (4) in \cite{b21}.)

\section{results}
\subsection{rms-flux relations in Proposal P10066}
\begin{figure}
\begin{center}
\includegraphics[width=3.3cm,angle=-90]{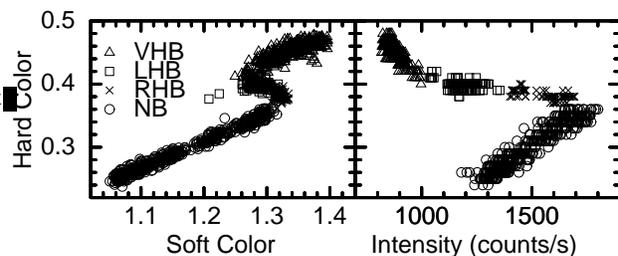}
\vspace{1mm} 
\caption{{\small The CCD (left panel) and HID (right panel) of Proposal P10066. 
The hard and soft color is defined as the flux ratio between 11.2-18.8keV and 
7.5-11.2kev, and between 4.7-7.5keV and 2-4.7keV, respectively. The Intensity is
defined as the count rate of 2-18.8keV.} } 
\label{ccd}
\end{center}
\end{figure}

\begin{figure*}
\vspace{1mm}
\begin{center}
\hspace{3mm}
\includegraphics[width=3.5cm,angle=270,clip]{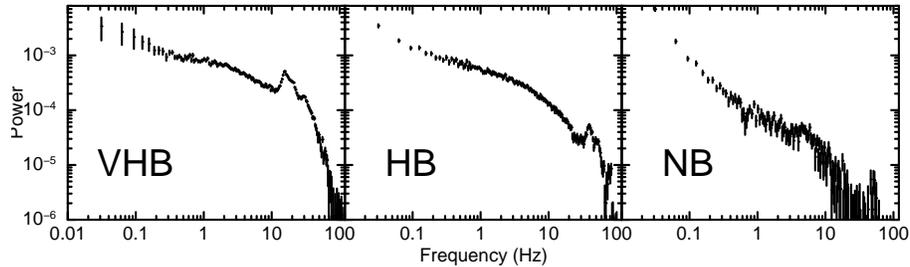}
\caption{{\small The PDSs of VHB(left), HB(middle) and NB(right) for Proposal P10066.}} 
\label{pds}
\end{center}
\end{figure*}

\begin{figure*}
\vspace{1mm}
\begin{center}
\hspace{3mm}
\includegraphics[width=3.5cm,angle=270,clip]{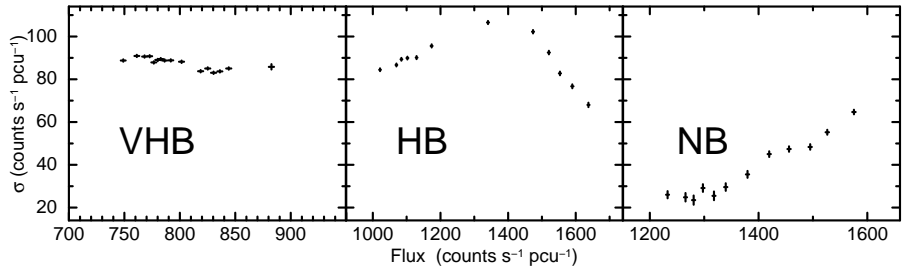}
\caption{{\small The rms-flux relations of VHB(left), HB(middle) and NB(right) for Proposal P10066}}
\label{rms-flux}
\end{center}
\end{figure*}

As a example, we will show the rms-flux relation of Proposal P10066
in this section, because it clearly shows three branches in the CCD.
Proposal P10066 includes three observations, i.e. 10066-01-01-000,
10066-01-01-001 and 10066-01-01-00. We display its CCD and hard
color-intensity diagram (HID) in Fig. \ref{ccd}. The CCD shows three
branches clearly, i.e., the top, middle and bottom branch. Their PDS
are shown in Fig. \ref{pds}. Because both the top and middle
branches exhibit QPOs between $\sim$15 to $\sim$60 Hz, they locate
in the HB. Given the shape of the top branch, we name it VHB, and
call the middle branch HB. The bottom branch is NB because of a power
law PDS and its $\sim$7 Hz QPO (see \cite{b7}). Using the method in
Section 2.2, we obtain the rms-flux relations of the three branches,
and the results are shown in Fig. \ref{rms-flux}.

\begin{figure}
\vspace{1mm}
\begin{center}
\hspace{3mm}
\includegraphics[width=3.5cm,angle=270,clip]{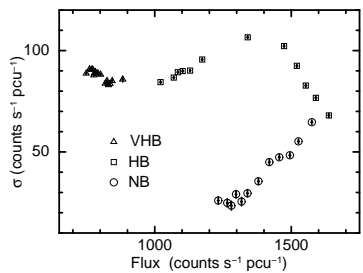}
\caption{{\small The complete rms-flux relation of Proposal P10066.}}
\label{10066_rms}
\end{center}
\end{figure}

Our analysis shows that the rms is negatively correlated with flux
in the VHB, but positively correlated in the NB of P10066 (Fig.
\ref{rms-flux}). In the HB of P10066, the rms-flux relation is not a
linear correlation anymore. In fact, the rms positively correlated
with flux in the left part of HB and negatively correlated in the
right. In this manner the rms has a maximum value at the flux of
$\sim$1400 cts/s, approximately corresponding to the middle part of
HB. For comparison we show the whole rms-flux relation of Proposal
P10066 in Fig. \ref{10066_rms}. By using a linear model to fit the
relation of each branch in Fig. \ref{10066_rms}, we derive the
turn-over points between different branches. In the VHB of P10066,
the flux is less than $\sim$950 cts/s/pcu and the QPO frequency of
each flux bin is less than $\sim$30 Hz. In the left HB of P10066, the
flux is more than $\sim$950 cts/pcu and less than $\sim$1430
cts/s/pcu, and its QPO frequency is between $\sim$30 Hz and
$\sim$50 Hz. The flux of the right HB of P10066 is more than
$\sim$1430cts/s/pcu and its QPO frequencies range from $\sim$50 Hz to
$\sim$55 Hz. Those turn-over points are listed in Table
\ref{turn-over}.

\subsection{rms-flux relations in the other observations}
\begin{figure*}
\begin{center}
\includegraphics[width=3.5cm,angle=270,clip]{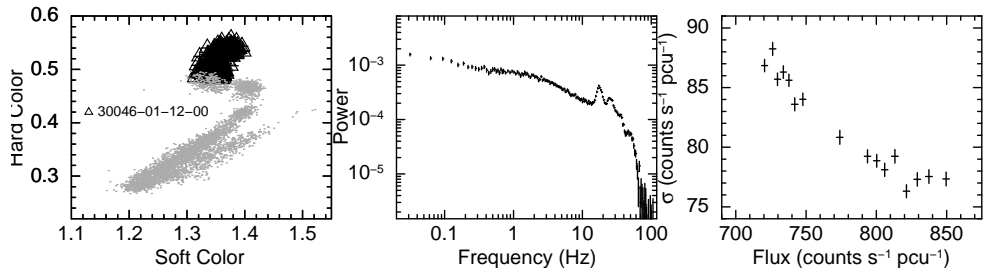}
\caption{{\small This figure shows the CCD(left), PDS(middle) and rms-flux
relation(right) of ObsID 30046-01-12-00, which locates at the VHB
and displays a negative correlation.}} 
\label{vhb}
\end{center}
\end{figure*}

\begin{figure*}
\begin{center}
\includegraphics[width=3.5cm,angle=270,clip]{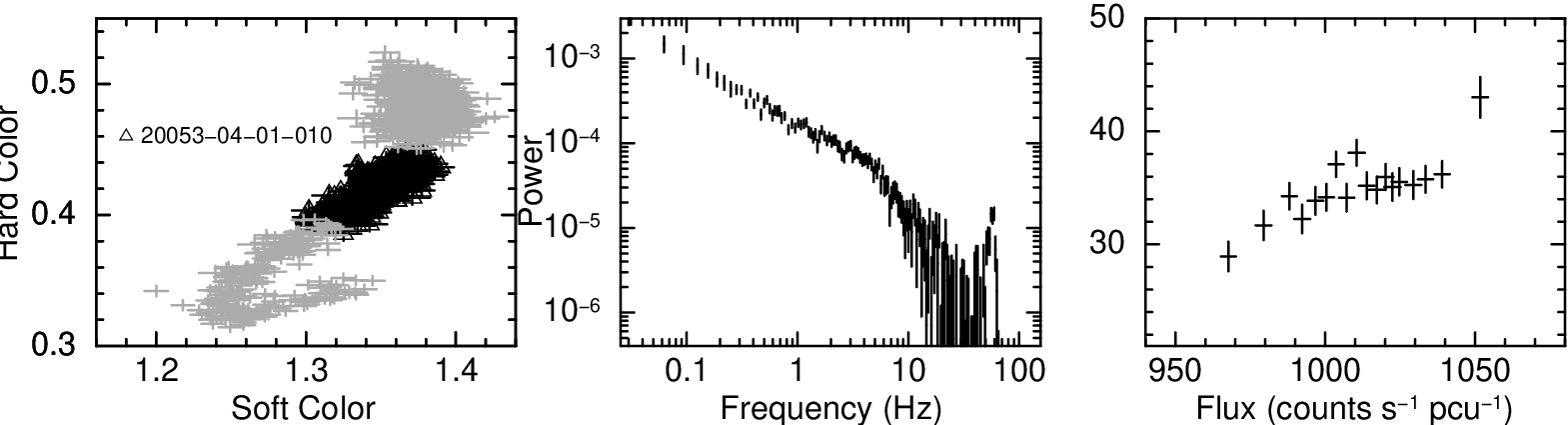}
\includegraphics[width=3.5cm,angle=270,clip]{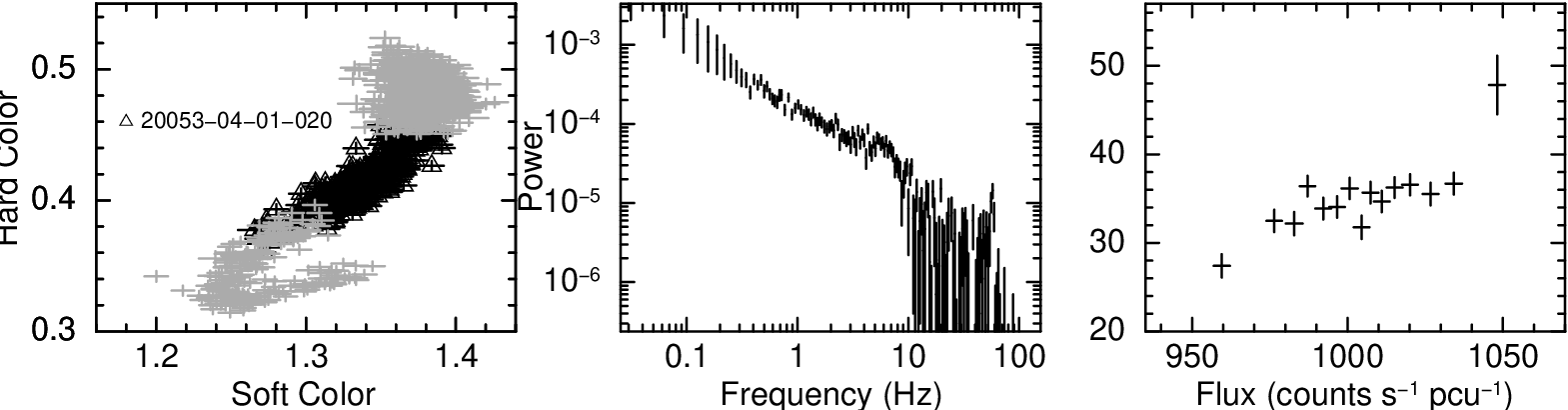}
\caption{{\small The CCDs (left column), PDSs (middle column) and rms-flux relations (right 
column) of observations locate in the NB. The top and bottom panels show the results of 
ObsIDs 20053-04-01-010 and 20053-04-01-010 respectively.}} 
\label{nb}
\end{center}
\end{figure*}

\begin{figure*}
\begin{center}
\includegraphics[width=3.5cm,angle=270,clip]{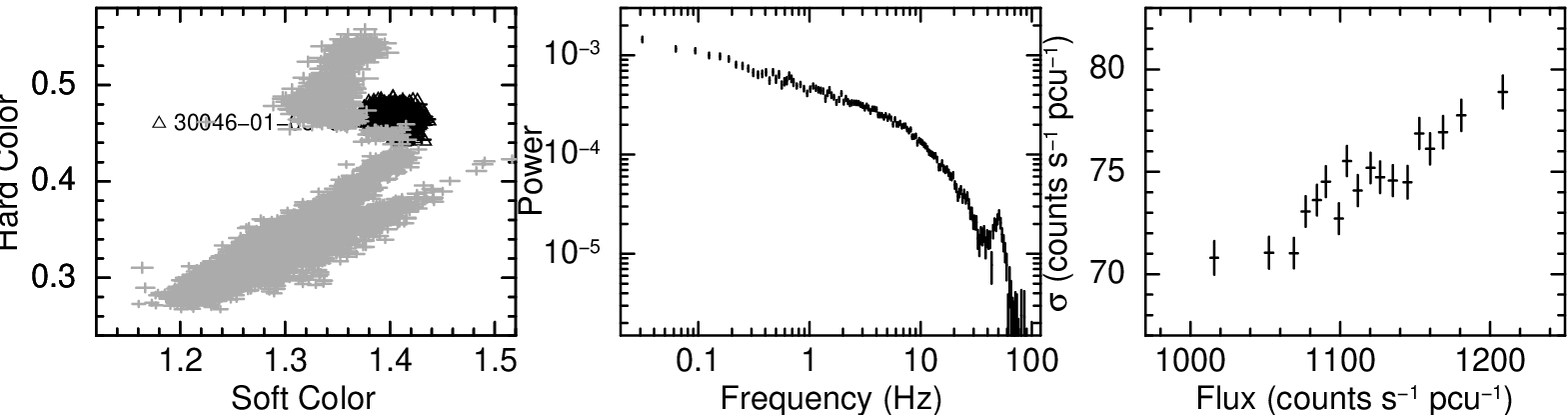}
\includegraphics[width=3.5cm,angle=270,clip]{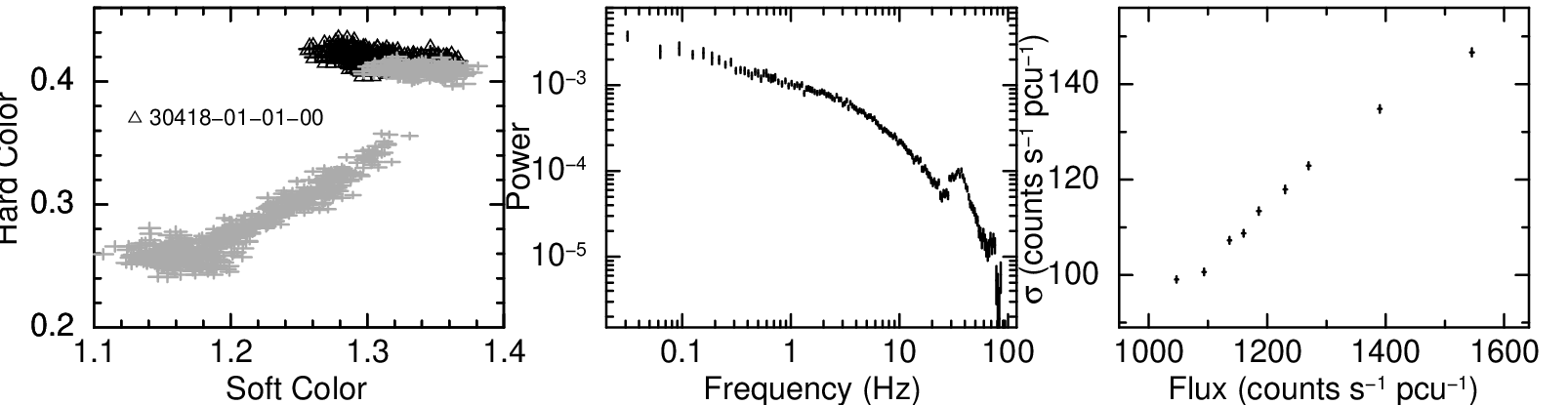}
\caption{{\small The CCDs (left column), PDSs (middle column) and rms-flux relations (right 
column) of observations locate in the left part of HB. The top and bottom panels show the 
results of ObsIDs 30046-01-03-00 and 30418-01-01-00 respectively.}} 
\label{lefthb}
\end{center}
\end{figure*}

\begin{figure*}
\begin{center}
\includegraphics[width=3.5cm,angle=270,clip]{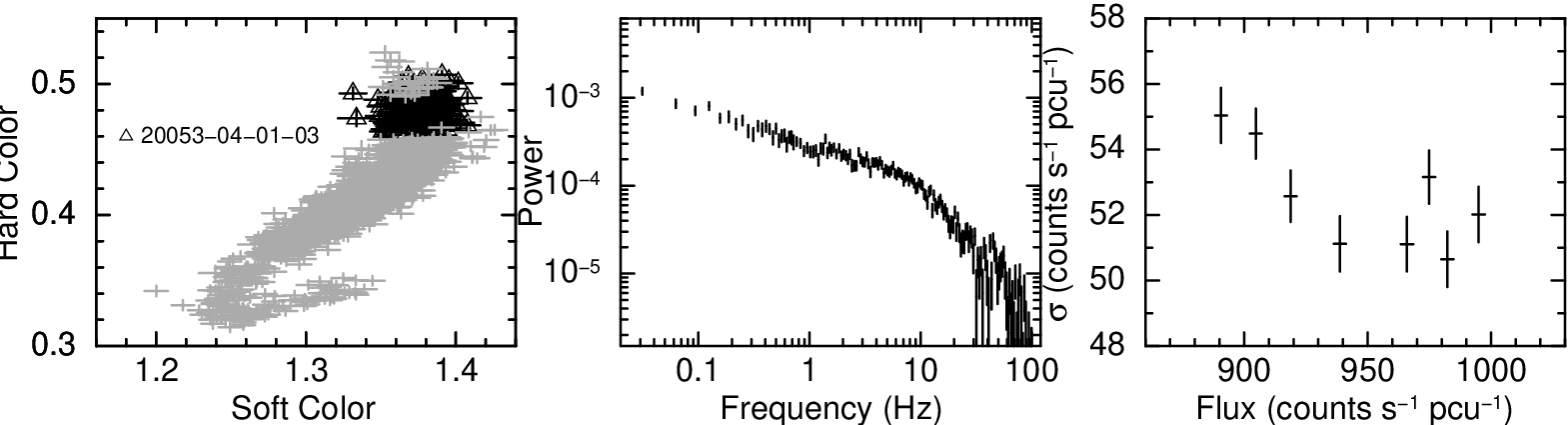}
\includegraphics[width=3.5cm,angle=270,clip]{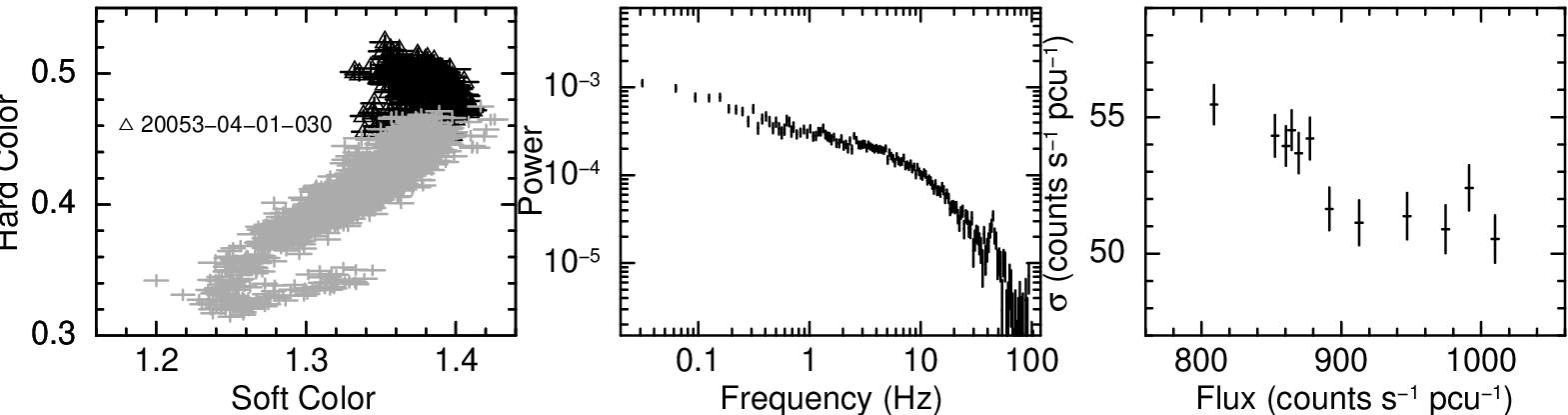}
\includegraphics[width=3.5cm,angle=270,clip]{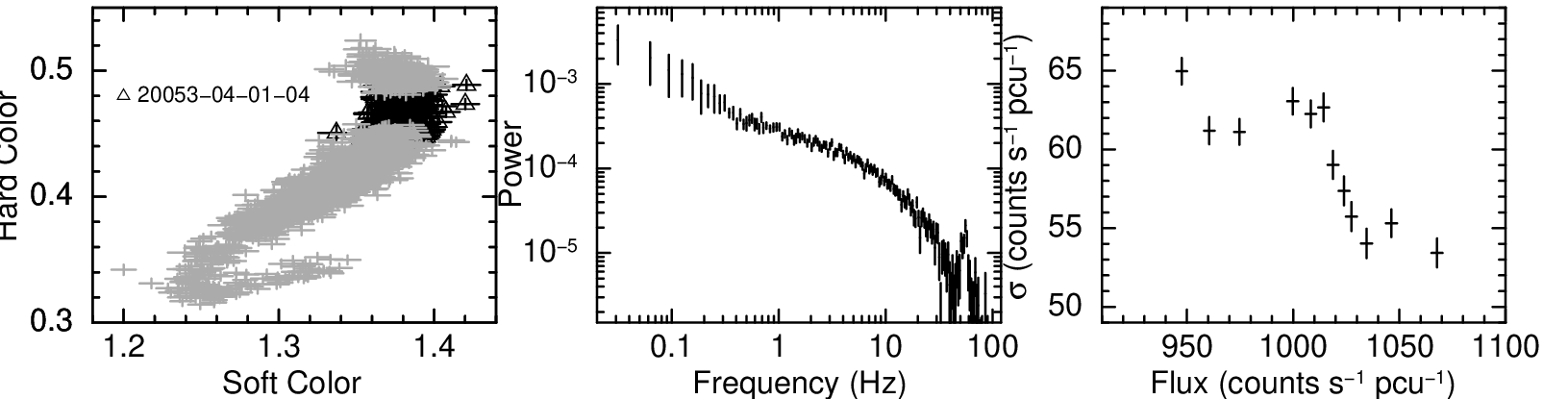}
\caption{{\small The CCDs (left column), PDSs (middle column) and rms-flux relations (right 
column) of observations locate in the right part of HB. The top, middle and bottom panels show 
the results of ObsIDs 20053-04-01-03, 20053-04-01-030 and 20053-04-01-04, respectively.}} 
\label{righthb}
\end{center}
\end{figure*}

We derive the rms-flux relations of all the other observations
listed in Table \ref{ObsIDs}. The results are shown in Fig.
\ref{vhb}-\ref{vertex}. Fig. \ref{vhb} show the results of ObsID
30046-01-12-00, who locates in the VHB of the CCD. Its flux bin 
frequencies and flux are respectively less
than $\sim$30 Hz and $\sim$950cts/s/pcu, and the rms-flux relation is
negative. These results are similar to those of the VHB P10066. Fig.
\ref{hb} shows the results of these observations locating in HB.
There rms-flux relations are the same as in the HB of P10066. By
fitting their rms-flux relations, we derive the turn-over points are
$\sim$1460cts/s/pcu, $\sim$1480cts/s/pcu, $\sim$1430cts/s/pcu and
$\sim$1400cts/s/pcu for 10065-02-01-000, 10065-02-01-001, 10065-02-01-002, 
and 10065-02-01-003, respectively. Moreover, their QPO frequencies
are between $\sim$38 Hz and $\sim$50 Hz for the left HB, and between
$\sim$50 Hz and $\sim$55 Hz for the right HB(Table \ref{turn-over}).
The ObsIDs 20053-04-01-010 and 20053-04-01-020 locate in the NB and
their results are shown in Fig \ref{nb}. There two observations do
not have obvious QPOs but a bump at $\sim$5-7 Hz, and their hard
colors are less than $\sim$0.45. Although their flux, between
$\sim$950cts/s/pcu and $\sim$1100cts/s/pcu, are less than in the NB
of P10066, their rms-flux relations are positive.

Fig. \ref{lefthb} shows the results of these observations locate
only in the left part of HB. For the ObsID 30046-01-03-00, its flux
is less than $\sim$1300cts/s/pcu and its frequency is between
$\sim$45 Hz and $\sim$53 Hz. However, the situation is a little
different for ObsID 300418-01-01-00. Its flux is less than
$\sim$1500cts/s/pcu and the QPO frequencies range from $\sim$30 Hz to
$\sim$48 Hz. As a result, their rms-flux relations are positive.
There are three observations locate in the right part of HB(see
Table \ref{ObsIDs}) and their results are shown in Fig.
\ref{righthb}. Their hard colors are more than $\sim$0.45 while
their flux are between $\sim$800cts/s/pcu and $\sim$1100cts/s/pcu.
For ObsIDs 20053-04-01-03 and 20053-04-01-04, the QPO frequencies
are around $\sim$53-55 Hz, but it locates around $\sim$45 Hz in the
ObsID 20053-04-01-030. The same as in the right HB of P10066, their
rms are negatively correlated with flux. Because these observations
shown in Fig. \ref{lefthb}-\ref{righthb} occupy only a portion of
HB, their rms-flux relations show only a portion of the arch-like
relation of the HB of P10066 either.

In addition, there is one observation, 30046-01-08-00, locating at
the vertex of the VHB and HB (see Fig. \ref{vertex}). This
observation has two branches. The flux of the VHB part is less than
$\sim$960 cts/s/pcu, and its QPO frequencies range from $\sim$26 Hz
to $\sim$35 Hz, while the flux of the left HB part is more than
$\sim$ 960 cts/s/pcu and its QPO frequencies are between $\sim$37 Hz
and $\sim$50 Hz(see Table \ref{turn-over}). As shown in the right
panel of Fig. \ref{vertex}, the VHB part shows a negative
correlation in the rms-flux plot, and the left HB part shows a
positive one, which are consistent with those results above.

\subsection{the correlations between rms and spectral parameters in P10066}
Because the other observations have the same results of rms-flux relations 
as in Proposal P10066, our spectral fitting concentrate only on Proposal
P10066. First, we split the data into 25 intervals according to its
position on the CCD. The average hard color (HC) and soft color (SC)
of the 25 intervals are listed in Table \ref{chi-squared}. 
(Because there is no need to correct the gain difference between
Epoches II and III, we haven't corrected the colors in this
subsection and they are different from above.) Second, we compute
their rms using the method described above (see Table
\ref{chi-squared}). Finaly, we extract their spectra with the
Standard 2 mode data and use the X-ray spectral-fitting program,
XSPEC, to analyze their spectra.

Considering the controversy over the location and nature of the
X-ray emission regions, we fit their spectra with the Eastern and
Western model, respectively. We use the absorbed disk blackbody,
Comptonization plus an extra Gaussian line (wabs(
diskbb+compTT+gauss)) as the Eastern model, and the absorbed blackbody,
Comptonization plus an extra Gaussian line (wabs(bb+compTT+gauss))
as the Western model, i.e. the Birmingham model \citet{b3}. The
fitting energy band is 3-30keV and the absorption column density
N$_h$ is fixed at 0.2$\times 10^{22}$. According to their reduced
$\chi^2$ values shown in Table \ref{chi-squared}, we can find that
the Western model gives a better fitting results for those intervals
whose ID $>$ 17. We also give the inner radius of the accretion disk
of the Eastern Model in Table \ref{chi-squared} by assuming an
inclination angle of the system of 60$^{o}$ (see \citep{b38}).
Because the inner disk radius is less than the typical radius of a
NS, i.e., $\sim$10km, the Eastern Model isn't reasonable for Cyg X-2
in this situation. The spectral parameters of the Birmingham model
are listed in Table \ref{parameters}. The relations between the rms
and flux of these two components of Birmingham model is shown in
Fig. \ref{parameter_rms}. We also present the relation between the
rms and flux ratio of blackbody and Comptonization components in
Fig. \ref{rms_ratio}.

\section{Discussion}

\begin{figure*}
\begin{center}
\includegraphics[width=3.5cm,angle=270,clip]{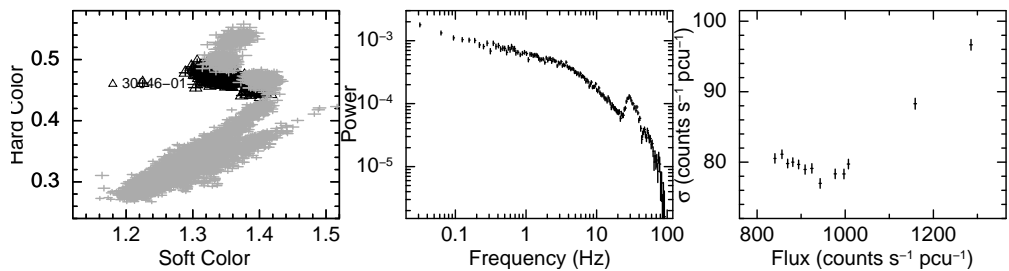}
\caption{{\small The CCD (left), PDS (middle) and rms-flux relation (right) of ObsID 
30046-01-08-00, which locates in the vertex of VHB and HB.}} 
\label{vertex}
\end{center}
\end{figure*}

\citet{b15} firstly reported a negative rms-flux relation in Cyg
X-2, and then it is also found in the black hole XRB XTE J1550-564
(see also \cite{b10}). The positive relation can, in principle, be
readily understood: the larger the flux, the higher the accretion
rate which leads to a higher level of instability. Nevertheless, the
negative correlation between rms and flux in Cyg X-2 and XTE
J1550-564 is hard to fit into this scenario (see \cite{b15,b10}).
As shown in Fig. \ref{vertex}, the complicated rms-flux relation of
Cyg X-2 leads to a maximal rms in the middle of HB, which means that
Cyg X-2 has a maximum level of instability no matter how luminous it
is. Moreover, the negative relation dose not change with the
integral frequency range. In other words, if we do not include the
QPO and only integral from 0.125 Hz to 10 Hz when we compute the rms,
the rms-flux relation is also negative in the VHB. This means that
the negative relation is not caused by the QPO. We also normalize
the rms by dividing the flux of each flux bin, and the normalized
rms shows a positive correlation with flux in the NB, but negative
in the VHB, which is totally the same as not normalized. The same
results mean that the rms-flux relation isn't due to the flux change
of different flux bins.

To compare NSXRB Cyg X-2 with BHXRB Cyg X-1, we investigate the
rms-flux relations in different branches in Cyg X-2. As shown above,
the rms is positively correlated with flux when Cyg X-2 locates in
the NB or at the left part of HB. However, it is negatively
correlated with flux in the VHB or at the right part of HB. This is
different from those reported in \citet{b26,b27,b9}. Especially,
\citet{b6} found that the positive rms-flux relation is valid
throughout all states of the BHXRB Cyg X-1. Moreover, the slope of
the rms-flux relation of the hard state is steeper than in the soft
and intermediate states of Cyg X-1. However, the slopes of the hard
state (VHB state and its slope is negative) is flatter than in the
soft state (NB state, its slope is positive) in Cyg X-2. We guess
that this difference may come from the compact star.

However, although Cyg X-1 and Cyg X-2 have different results in the
hard and soft states, the rms-flux relation shows some similarities
if we assume HB is the intermediate state of Cyg X-2. \citet{b6}
reported an 'arch'-like rms-flux relation in the intermediate state
of Cyg X-1, but they claimed that the 'arch' relation is due to the
buffer overflow.

\begin{figure*}
\begin{center}
\includegraphics[width=3.5cm,angle=270,clip]{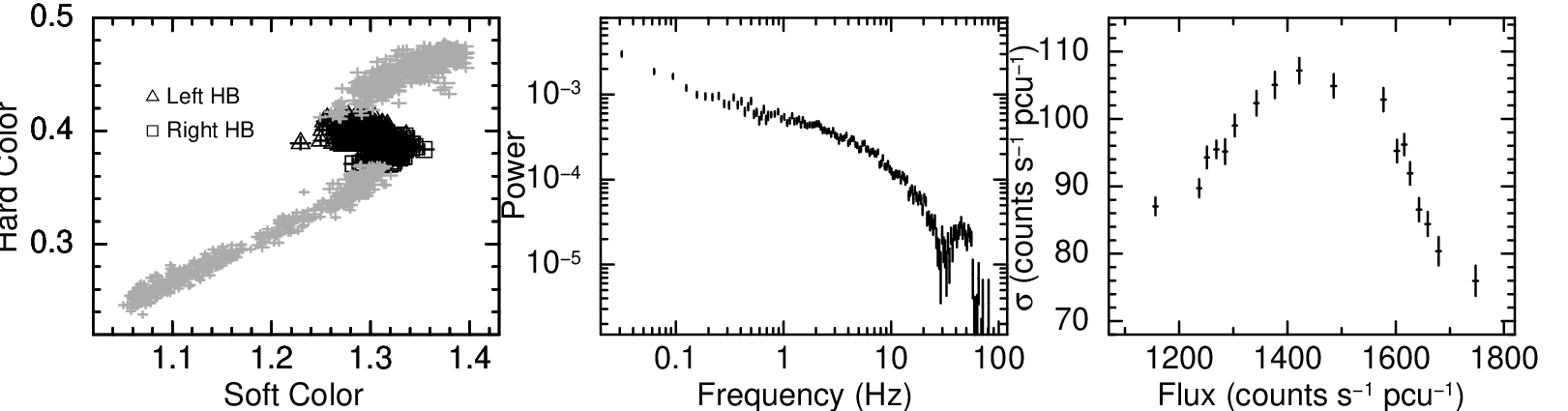}
\includegraphics[width=3.5cm,angle=270,clip]{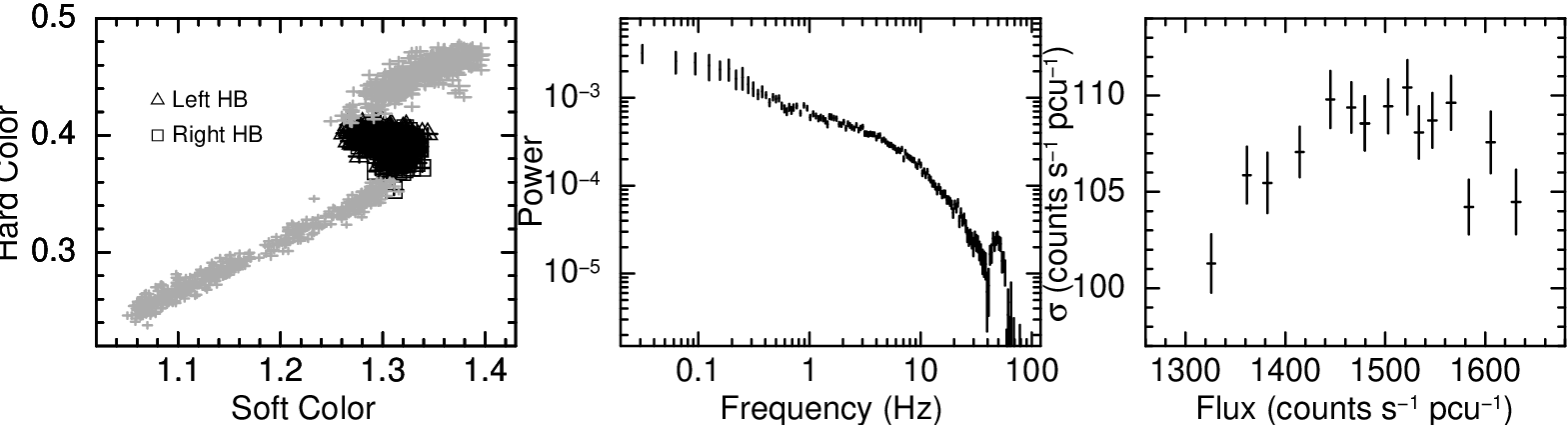}
\includegraphics[width=3.5cm,angle=270,clip]{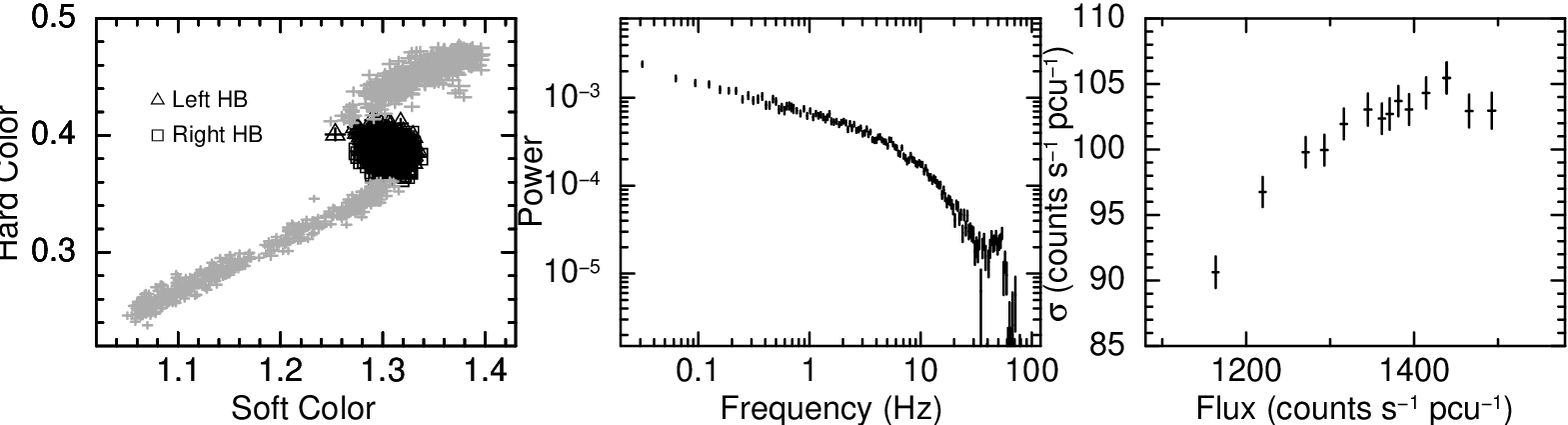}
\includegraphics[width=3.5cm,angle=270,clip]{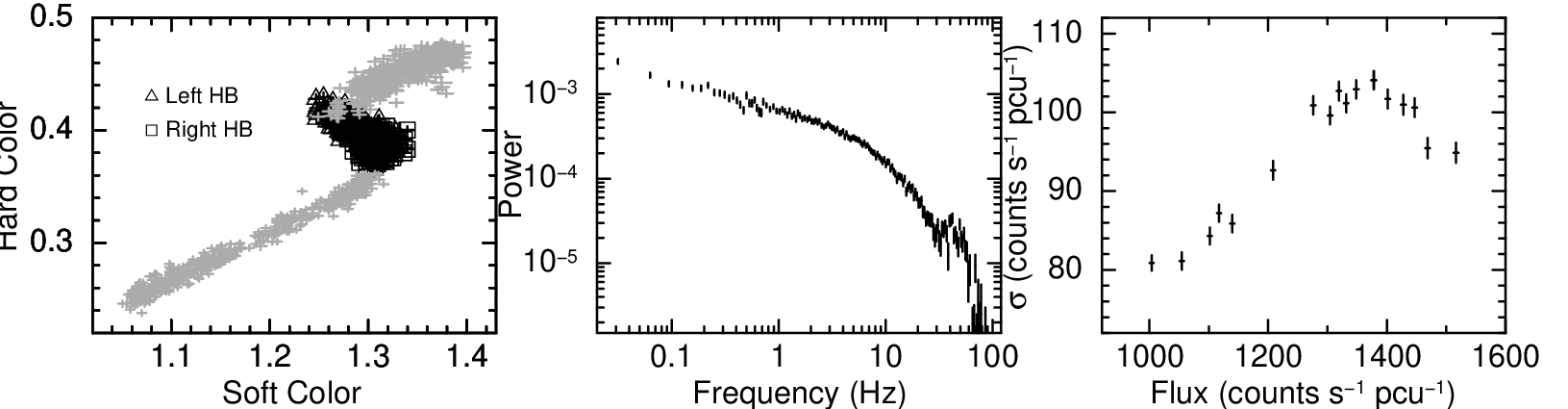}
\caption{{\small The CCDs (left column), PDSs (middle column) and rms-flux relations (right 
column) of those observations locating at the HB. Row 1, 2, 3 and 4, show the relults of 
ObsIDs 10065-02-01-000, 10065-02-01-001, 10065-02-01-002 and 10065-02-01-003, respectively. 
The left HB is marked with triangles and the right HB with squares in each CCD.}} 
\label{hb}
\end{center}
\end{figure*}

The so-called buffer overflow occurs when a program or process tries
to store more data in a buffer (temporary data storage area) than it
could hold. Due to the softness of the X-ray spectrum, the
likelihood for such overflows is the highest in the lowest energy
band. Still, for the different configuration, the possibility of
overflow is lower when the configuration threshold is higher. In
this work, we also find some 'arch' relations in the intermediate
state of Cyg X-2 and try to determine wether the arch shape in Cyg
X-2 is due to the buffer overflow or not. We've tried the Event
Encoded mode data (see Table \ref{ObsIDs} and Fig. \ref{hb}), the
higher threshold mode data, and higher energy band data, which are
less likely suffered from buffer overflows (also see
http://heasarc.gsfc.nasa.gov/docs/xte/RXTE\_tech\_ap\\
pend.pdf)), and
found that these data also show similar arch relations. On the other
hand, we've done some simulations and the simulation results don't
support the argue that the arch relation is due to the buffer
overflow. As a result, we conclude that the 'arch' rms-flux relation
may be the characteristic of the state transition, which always take
place at the QPO frequency $\sim$50 Hz(see Table \ref{turn-over}).

In the 1980s, two models were proposed to explain the X-ray emission
of the NSXRB: the Western model and the Eastern model. The Eastern
model contains a multi-temperature blackbody from the inner disc,
plus a Comptonized emission of the seed photons from the blackbody
of the NS \citep{b19}, while the Western model contains
a Comptonization component and sometimes a second one (i.e.,
blackbody emission) for some luminous sources \citep{b33}. In the
1990s, a new model - the Birmingham model, closely related to the
Western model - is proposed. This model assumes that all LMXBs have
two continuum components: blackbody emission from the NS surface and
Comptonized emission from an extended accretion disc corona above
the accretion disc \citep{b3}. Our spectral results show that the
Birmingham model is more reasonable than the Eastern model (see
Table \ref{chi-squared}), and this suggests that the two components
in Cyg X-2 are most likely the blackbody coming from the surface of
the NS and the Comptonization generated from the corona.

\begin{figure}
\begin{center}
\includegraphics[width=3cm,angle=270,clip]{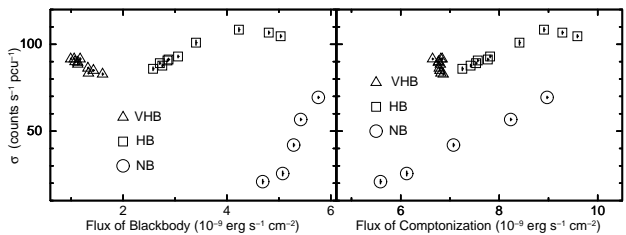}
\caption{{\small The rms-flux plot. The X-axis of the left panel is the flux of blackbody, and 
the right is the flux of Comptonization component.}} 
\label{parameter_rms}
\end{center}
\end{figure}

From Table \ref{parameters} we know that the temperatures of the
seed photons, the blackbody and the corona, the optical depth, and
the flux of the Comptonization component change little, while the
blackbody flux increases apparently in the VHB. A scenario hinted
from these results may be that the corona is not efficiently cooled
by the soft disk emission, and the latter is also likely stable
because of the little change in T0. As a result most material move
onto the surface of NS, and the rms changes obviously. In other
words, the negative rms-flux relation of the VHB is supposed to
contribute from the blackbody component (Fig. \ref{parameter_rms}).
Similarly, the rms is dominated by the Comptonization component in
the NB (Fig. \ref{parameter_rms}). To sum up, the blackbody
component dominates the rms variability in VHB and its contribution
to rms decreases as Cyg X-2 evolving through the HB to NB, whereas
the rms contributed by the Comptonization component increases and
dominates the rms variability in NB at last. This inference is also
supported by the results that the rms negatively correlated with
flux in energy bands 2-6keV and 6-10keV, and positively correlated
in 10-21keV (see \cite{b15}). Because of these totally different
manners of these two components, Cyg X-2 has a complex rms-flux
relation along with the CCD track.

\begin{figure}
\begin{center}
\includegraphics[width=3.5cm,angle=270,clip]{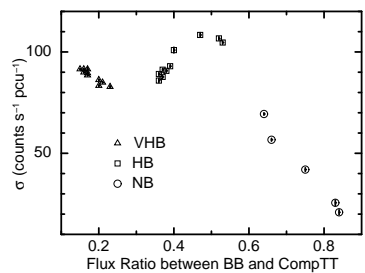}
\caption{{\small The rms-ratio plot of Proposal P10066. The X-axis stands for its flux ratio 
between blackbody and Comptonization components.}} 
\label{rms_ratio}
\end{center}
\end{figure}

Fig. \ref{rms_ratio} shows that the instability of Cyg X-2 is
related to the flux ratio between the blackbody and Comptonization
components. Only when those two components reach a certain ratio,
Cyg X-2 reaches its maximum instability. This can be explained in
the Birmingham model. In this model, the radiation of the blackbody
shines on the accretion disc and forms a corona. On the one hand,
the rms produced by the blackbody decreases when the flux of the
blackbody increases. On the other hand, the radiation of the
increasing blackbody will expand the corona region and increase the
flux of the Comptonization component, which will finally increase
its rms. As a result, the rms has a maximum only when their flux
reaches a certain ratio.

\acknowledgments This work is subsidized by the Program of the Light
in Chinese Western Region (LCWR) (Grant No. XBBS201121) provided by
Chinese Academy of Sciences (CAS), the Natural Science Foundation of
China for support via NSFC 11173034, 10903005, 11173024, 10473010
and 19673010, CAS key project via KJCX2-YWT03 and National Basic
Research Program of China (2009CB824800). We thank S. N. Zhang, H.
Tong, F. J. Lu, Y. P. Chen, S. J. Zheng, and M. Y. Ge for useful
discussions.


%
\bibliography{template}                
\bibliographystyle{spr-mp-nameyear-cnd}  

%

\begin{table*}[]
\tabcolsep 1mm 
\caption[]{{\small The information of those observations used in this work.}}
\label{ObsIDs}
\begin{center}
\begin{tabular}{cccccc}
\hline\noalign{\smallskip}
ObsID  & date & epoch & duration  &   data mode  &   Branch     \\
\hline\hline\noalign{\smallskip}
10065-02-01-000 & 23/3/96 & II & 14ks & E\_125us\_64M\_0\_1s  & HB \\
10065-02-01-001 & 23/3/96 & II & 25ks & E\_125us\_64M\_0\_1s  & HB \\
10065-02-01-002 & 23/3/96 & II & 19ks & B\_2ms\_16A\_0\_35\_Q & HB \\
10065-02-01-003 & 23/3/96 & II & 15ks & B\_2ms\_16A\_0\_35\_Q & HB \\
10066-01-01-00  & 27/3/96 & II & 16ks & B\_2ms\_16A\_0\_35\_Q & VHB \\
10066-01-01-000 & 26/3/96 & II & 16ks & B\_2ms\_16A\_0\_35\_Q & NB \\
10066-01-01-001 & 27/3/96 & II & 18ks & B\_2ms\_16A\_0\_35\_Q & HB \\
20053-04-01-010 & 1/7/97 & III & 25ks & B\_2ms\_16A\_0\_35\_Q & NB \\
20053-04-01-020 & 1/7/97 & III & 15ks & B\_2ms\_16A\_0\_35\_Q & NB \\
20053-04-01-03  & 2/7/97 & III & 12ks & B\_2ms\_16A\_0\_35\_Q & right HB \\
20053-04-01-030 & 2/7/97 & III & 13ks & B\_2ms\_16A\_0\_35\_Q & right HB \\
20053-04-01-04  & 2/7/97 & III & 15ks & B\_2ms\_16A\_0\_35\_Q & right HB \\
30046-01-03-00  & 28/7/98 & III & 15ks & B\_2ms\_16A\_0\_35\_Q & left HB \\
30046-01-08-00  & 30/8/98 & III & 12ks & B\_2ms\_16A\_0\_35\_Q & VHB-HB vertex \\
30046-01-12-00  & 25/9/98 & III & 20ks & B\_2ms\_16A\_0\_35\_Q & VHB \\
30418-01-01-00  & 2/7/98 & III & 12ks & B\_2ms\_16A\_0\_35\_Q & left HB \\
\hline\noalign{\smallskip}
\end{tabular}
\end{center}
\end{table*}

\begin{table*}[]
\tabcolsep 1mm 
\caption[]{{\small The turn-over points between different
branches. R and f stand for the flux in the unit of cts/s/pcu and the frequency of the QPO or
the bump in the unit of Hz, respectively.}} 
\label{turn-over}
\begin{center}
\begin{tabular}{cccccc}
\hline\noalign{\smallskip}
ObsID  & VHB & left HB & right HB  &   NB     \\
\hline\hline\noalign{\smallskip}
10065-02-01-000 & --- & R$<$1460, 38$<$f$<$50 & 1460$<$R, 50$<$f$<$55 & --- \\
10065-02-01-001 & --- & R$<$1480, 38$<$f$<$50 & 1480$<$R, 50$<$f$<$55 & --- \\
10065-02-01-002 & --- & R$<$1430, 38$<$f$<$50 & 1430$<$R, 50$<$f$<$55 & --- \\
10065-02-01-003 & --- & R$<$1400, 38$<$f$<$50 & 1400$<$R, 50$<$f$<$55 & --- \\
VHB of P10066  & R$<$950, f$<$30 & --- & --- & --- \\
left HB of P10066  & --- & 950$<$R$<$1430, 30$<$f$<$50 & --- & --- \\
right HB of P10066 & --- & --- & 1430$<$R, 50$<$f$<$55 & --- \\
NB of P10066  & --- & --- & --- & 5$<$f$<$7 \\
20053-04-01-010 & --- & --- & --- & 5$<$f$<$7 \\
20053-04-01-020 & --- & --- & --- & 5$<$f$<$7 \\
20053-04-01-03  & --- & --- & 800$<$R$<$1100, f$\sim$53-55 & --- \\
20053-04-01-030 & --- & --- & 800$<$R$<$1100, f$\sim$53-55 & --- \\
20053-04-01-04  & --- & --- & 800$<$R$<$1100, f$\sim$45 & --- \\
30046-01-03-00  & --- & R$<$1300, 45$<$f$<$53 & --- & --- \\
30046-01-08-00  & R$<$960, 26$<$f$<$35 & 960$<$R, 37$<$f$<$50 & --- & --- \\
30046-01-12-00  & R$<$950, f$<$30 & --- & --- & --- & --- \\
30418-01-01-00  & --- & R$<$1500, 30$<$f$<$48 & --- & --- & --- \\
\hline\noalign{\smallskip}
\end{tabular}
\end{center}
\end{table*}

\begin{table}[]
\tabcolsep 1mm 
\caption[]{{\small The $\chi^2$ of the Western and Eastern models. R$_{in}$ is 
the inner disk radius derived from the Eastern Model by assuming an inclination 
angle of the system of 60$^{o}$.}} 
\label{chi-squared}
\begin{center}
\begin{tabular}{ccccccc}
\hline\noalign{\smallskip}
ID & HC & SC & Western$^{1}$ & Eastern$^{2}$ & rms  & R$_{in}$$^{3}$ \\
\hline\noalign{\smallskip}
1 & 0.395 & 0.885 & 0.92(56) & 0.91(56) & 91.5$\pm$0.4 & 3.3 \\
2 & 0.394 & 0.885 & 0.83(56) & 0.89(56) & 91.5$\pm$0.4 & 4.2 \\
3 & 0.391 & 0.883 & 0.89(56) & 0.91(56) & 90.0$\pm$0.4 & 4.6  \\
4 & 0.389 & 0.875 & 1.18(56) & 1.24(56) & 89.6$\pm$0.4 & 4.3  \\
5 & 0.389 & 0.876 & 1.06(56) & 1.07(56) & 88.5$\pm$0.4 & 4.7  \\
6 & 0.386 & 0.872 & 0.80(56) & 0.82(56) & 91.5$\pm$0.7 & 4.4  \\
7 & 0.377 & 0.856 & 0.55(56) & 0.56(56) & 86.2$\pm$0.4 & 5.8  \\
8 & 0.373 & 0.850 & 0.62(56) & 0.65(56) & 83.4$\pm$0.4 & 6.1  \\
9 & 0.372 & 0.848 & 0.96(56) & 0.99(56) & 85.0$\pm$0.4 & 6.6  \\
10 & 0.364 & 0.839 & 1.00(56) & 1.04(56) & 82.8$\pm$0.6 & 6.3  \\
11 & 0.343 & 0.830 & 0.82(56) & 0.86(56) & 85.9$\pm$1.0 & 8.3  \\
12 & 0.340 & 0.832 & 0.57(56) & 0.65(56) & 87.7$\pm$0.7 & 8.5  \\
13 & 0.339 & 0.830 & 0.74(56) & 0.86(56) & 89.1$\pm$0.7 & 9.1  \\
14 & 0.339 & 0.833 & 0.69(56) & 0.83(56) & 90.7$\pm$0.7 & 8.9  \\
15 & 0.338 & 0.839 & 0.65(56) & 0.77(56) & 91.2$\pm$0.7 & 9.0  \\
16 & 0.338 & 0.840 & 0.76(56) & 1.11(57) & 93.0$\pm$0.7 & 8.9  \\
17 & 0.334 & 0.850 & 0.73(56) & 1.41(57) & 100.9$\pm$1.4 & 8.8  \\
18 & 0.329 & 0.857 & 0.98(56) & 1.55(57) & 108.4$\pm$1.1 & 8.9  \\
19 & 0.321 & 0.858 & 0.62(56) & 1.37(57) & 106.7$\pm$0.9 & 8.9  \\
20 & 0.322 & 0.861 & 0.47(56) & 1.42(57) & 104.7$\pm$0.8 & 9.1  \\
21 & 0.295 & 0.835 & 0.73(56) & 1.57(57) & 69.4$\pm$0.6 & 10.6  \\
22 & 0.282 & 0.812 & 0.62(56) & 1.61(57) & 56.7$\pm$0.7 & 11.3  \\
23 & 0.258 & 0.768 & 0.77(56) & 1.72(57) & 42.0$\pm$0.9 & 12.4  \\
24 & 0.233 & 0.726 & 1.06(56) & 2.27(57) & 25.6$\pm$1.4 & 13.9  \\
25 & 0.220 & 0.702 & 1.61(56) & 3.63(57) & 20.9$\pm$1.2 & 15.2  \\
\hline\noalign{\smallskip}
\end{tabular}
\begin{list}{}{}
\item[$^{1}$]\small Western Model, i.e., wabs(bb+compTT+gauss).
\item[$^{2}$]\small Eastern Model, i.e., wabs(diskbb+compTT+gauss).
\item[$^{3}$]\small R$_{in}$ is derived from the Eastern Model in unit of km.
\end{list}
\end{center}
\end{table}

\begin{table*}[]
\tabcolsep 1mm 
\caption[]{{\small The fitting parameters of the Birmingham model. The fitting energy 
band is 3-30keV and the absorption column density N$_{h}$ is fixed at 0.2$\times 10^{22}$.} }
\label{parameters}
\begin{center}
\begin{tabular}{ccccccccc}
\hline\noalign{\smallskip}
ID & T0$^{1}$ & Tbb$^{1}$ & kT$^{1}$ & $\tau^{2}$ & Fe$^{3}$ & flux$_{bb}^{4}$ & flux$_{compTT}^{4}$ & flux$_{total}^{4}$ \\
\hline\noalign{\smallskip}
1 & 0.183$\pm$0.005 & 1.39$\pm$0.07 & 3.11$\pm$0.05 & 5.67$\pm$0.16 & 6.44$\pm$0.09 & 1.06 & 6.65 & 7.84$\pm$0.01 \\
2 & 0.194$\pm$0.004 & 1.27$\pm$0.05 & 3.02$\pm$0.04 & 5.94$\pm$0.12 & 6.46$\pm$0.09 & 0.99 & 6.81 & 7.96$\pm$0.01 \\
3 & 0.202$\pm$0.004 & 1.26$\pm$0.05 & 3.03$\pm$0.04 & 5.91$\pm$0.12 & 6.49$\pm$0.08 & 1.07 & 6.77 & 8.00$\pm$0.01 \\
4 & 0.186$\pm$0.004 & 1.26$\pm$0.05 & 3.02$\pm$0.04 & 5.89$\pm$0.13 & 6.58$\pm$0.08 & 1.13 & 6.76 & 8.02$\pm$0.01 \\
5 & 0.254$\pm$0.003 & 1.26$\pm$0.04 & 3.01$\pm$0.04 & 5.90$\pm$0.12 & 6.50$\pm$0.08 & 1.13 & 6.80 & 8.07$\pm$0.01 \\
6 & 0.202$\pm$0.004 & 1.29$\pm$0.05 & 3.02$\pm$0.05 & 5.80$\pm$0.14 & 6.48$\pm$0.08 & 1.17 & 6.84 & 8.13$\pm$0.01 \\
7 & 0.253$\pm$0.003 & 1.21$\pm$0.03 & 2.96$\pm$0.03 & 5.91$\pm$0.11 & 6.56$\pm$0.09 & 1.32 & 6.78 & 8.23$\pm$0.01 \\
8 & 0.215$\pm$0.003 & 1.19$\pm$0.03 & 2.94$\pm$0.03 & 5.89$\pm$0.10 & 6.49$\pm$0.09 & 1.33 & 6.82 & 8.28$\pm$0.01 \\
9 & 0.178$\pm$0.002 & 1.17$\pm$0.03 & 2.93$\pm$0.02 & 5.95$\pm$0.10 & 6.59$\pm$0.08 & 1.43 & 6.80 & 8.35$\pm$0.01 \\
10 & 0.214$\pm$0.003 & 1.21$\pm$0.03 & 2.95$\pm$0.03 & 5.77$\pm$0.11 & 6.50$\pm$0.10 & 1.60 & 6.86 & 8.57$\pm$0.01 \\
11 & 0.192$\pm$0.004 & 1.18$\pm$0.02 & 2.85$\pm$0.04 & 6.02$\pm$0.14 & 6.56$\pm$0.11 & 2.58 & 7.25 & 9.90$\pm$0.01 \\
12 & 0.177$\pm$0.004 & 1.17$\pm$0.02 & 2.85$\pm$0.04 & 6.04$\pm$0.13 & 6.59$\pm$0.14 & 2.75 & 7.42 & 10.27$\pm$0.01 \\
13 & 0.184$\pm$0.003 & 1.11$\pm$0.02 & 2.73$\pm$0.03 & 6.47$\pm$0.12 & 6.53$\pm$0.16 & 2.71 & 7.53 & 10.37$\pm$0.02 \\
14 & 0.219$\pm$0.003 & 1.14$\pm$0.02 & 2.78$\pm$0.03 & 6.31$\pm$0.12 & 6.62$\pm$0.12 & 2.85 & 7.57 & 10.53$\pm$0.02 \\
15 & 0.228$\pm$0.003 & 1.14$\pm$0.02 & 2.76$\pm$0.03 & 6.37$\pm$0.12 & 6.57$\pm$0.14 & 2.88 & 7.77 & 10.77$\pm$0.01 \\
16 & 0.185$\pm$0.003 & 1.12$\pm$0.01 & 2.73$\pm$0.03 & 6.60$\pm$0.12 & 6.60$\pm$0.14 & 3.06 & 7.81 & 11.01$\pm$0.01 \\
17 & 0.204$\pm$0.004 & 1.13$\pm$0.02 & 2.70$\pm$0.03 & 6.72$\pm$0.13 & 6.33$\pm$0.18 & 3.40 & 8.41 & 11.98$\pm$0.02 \\
18 & 0.207$\pm$0.005 & 1.17$\pm$0.02 & 2.70$\pm$0.03 & 6.76$\pm$0.15 & 6.58$\pm$0.15 & 4.23 & 8.91 & 13.22$\pm$0.02 \\
19 & 0.182$\pm$0.005 & 1.21$\pm$0.02 & 2.75$\pm$0.04 & 6.44$\pm$0.16 & 6.64$\pm$0.16 & 4.80 & 9.28 & 14.16$\pm$0.02 \\
20 & 0.184$\pm$0.005 & 1.18$\pm$0.01 & 2.68$\pm$0.03 & 6.86$\pm$0.15 & 6.59$\pm$0.17 & 5.03 & 9.59 & 14.70$\pm$0.02 \\
21 & 0.181$\pm$0.004 & 1.19$\pm$0.01 & 2.69$\pm$0.03 & 6.33$\pm$0.11 & 6.64$\pm$0.18 & 5.76 & 8.97 & 14.80$\pm$0.02 \\
22 & 0.190$\pm$0.004 & 1.16$\pm$0.01 & 2.63$\pm$0.03 & 6.24$\pm$0.11 & 6.62$\pm$0.11 & 5.42 & 8.23 & 13.74$\pm$0.02 \\
23 & 0.184$\pm$0.003 & 1.12$\pm$0.01 & 2.56$\pm$0.03 & 6.02$\pm$0.11 & 6.62$\pm$0.13 & 5.28 & 7.07 & 12.46$\pm$0.02 \\
24 & 0.196$\pm$0.003 & 1.07$\pm$0.01 & 2.41$\pm$0.02 & 6.03$\pm$0.10 & 6.57$\pm$0.09 & 5.07 & 6.12 & 11.32$\pm$0.01 \\
25 & 0.214$\pm$0.004 & 1.00$\pm$0.01 & 2.26$\pm$0.02 & 6.49$\pm$0.10 & 6.30$\pm$0.07 & 4.69 & 5.59 & 10.55$\pm$0.02 \\
\hline\noalign{\smallskip}
\end{tabular}
\begin{list}{}{}
\item[$^{1}$]\small Their units are keV and represent the temperature of seed photons, blackbody and the corona respectively.
\item[$^{2}$]\small $\tau$ stands for the optical depth. \\
\item[$^{3}$]\small Fe denotes the energy of the iron line. \\
\item[$^{4}$]\small Their units are $10^{-9} ergs s^{-1} cm^{-2}$ and represent the 3-30keV flux of blackbody, Comptonization component and total flux respectively.
\end{list}
\end{center}
\end{table*}

\end{document}